\documentclass[10pt,journal,comsoc]{IEEEtran}

\usepackage{amsmath,amsfonts}
\usepackage{algorithmic}
\usepackage{algorithm}
\usepackage{array}
\usepackage[caption=false,font=normalsize,labelfont=sf,textfont=sf]{subfig}
\usepackage{textcomp}
\usepackage{stfloats}
\usepackage{url}
\usepackage{graphicx}
\usepackage{cite}
\usepackage{xcolor}
\usepackage{hyperref}
\usepackage{booktabs}
\usepackage{balance}
\usepackage{enumitem}
\usepackage{fancyhdr}
\usepackage{etoolbox}

\setlist[itemize]{leftmargin=*,noitemsep,topsep=0pt}

\makeatletter
\def\footnoterule{\kern-3\p@\hrule\@width\z@\kern2.6\p@}
\makeatother

\title{Embodied AI in 6G Networks: From Intelligent Connectivity to Physical Intelligence
}

\author{%
  Junaid~Sajid,
  Sheikh~Salman~Hassan,~\IEEEmembership{Member,~IEEE,} Wenshuai~Liu,  
  Yan~Kyaw~Tun,~\IEEEmembership{Senior Member,~IEEE,} Yaru~Fu,~\IEEEmembership{Member,~IEEE,}~Nguyen~H.~Tran,~\IEEEmembership{Senior Member,~IEEE,}~Zhu~Han,~\IEEEmembership{Fellow,~IEEE,}\\
  Cedomir Stefanovic,~\IEEEmembership{Senior Member,~IEEE,}~Tharmalingam Ratnarajah, \IEEEmembership{Senior Member,~IEEE}, and Muhammad~Mahtab~Alam,~\IEEEmembership{Senior Member,~IEEE}%
  \thanks{%
    Junaid Sajid and Muhammad Mahtab Alam are with the Thomas Johann Seebeck Department of Electronics, Tallinn University of Technology (TalTech), 19086 Tallinn, Estonia
    Emails: junaid.sajid@taltech.ee; muhammad.alam@taltech.ee.
  }%
  \thanks{%
    Sheikh Salman Hassan is with the Institute for Imaging, Data and Communication (IDCoM), The University of Edinburgh, Edinburgh EH9 3BF, United Kingdom. Email: shassan@ed.ac.uk.
  }%
  \thanks{%
    Wenshuai Liu is with the School of Artificial Intelligence and Computer Science, Jiangnan University, Wuxi, Jiangsu, 214122, China, and also with the School of Science and Technology, Hong Kong Metropolitan University, Hong Kong, 999077, China. Email: liuws@stu.jiangnan.edu.cn.
  }%
  \thanks{%
    Yan Kyaw Tun and Cedomir Stefanovic are with the Department of Electronic Systems,
    Aalborg University, 9220 Aalborg, Denmark. Emails: ykt@es.aau.dk; cs@es.aau.dk.
  }%
  \thanks{%
    Yaru Fu is with the School of Science and Technology,  Hong Kong Metropolitan University, Hong Kong, 999077, China.
    Email: yfu@hkmu.edu.hk.
  }%
  \thanks{%
  Nguyen H. Tran is with the School of Computer Science, The University of Sydney, Australia. Email: nguyen.tran@sydney.edu.au.
  }%
  \thanks{%
  Zhu Han is with the Department of Electrical and Computer Engineering at the University of Houston, Houston, TX 77004 USA (e-mail: hanzhu22@gmail.com). }%
  \thanks{Tharmalingam Ratnarajah is with the Department of Electrical and Computer Engineering, San Diego State University, San Diego, CA 92182, USA. Email: t.ratnarajah@ieee.org.}
  \vspace{-0.3in}
  }

\begin{document}


\pagestyle{fancy}
\fancyhf{}

\renewcommand{\headrulewidth}{0pt}

\fancyhead[C]{\footnotesize
This work has been submitted to the IEEE Network Magazine for possible publication. Copyright may be transferred without notice, after which this version may no longer be accessible.
}

\fancyfoot[C]{\thepage}

\setlength{\headheight}{24pt}

\maketitle
\thispagestyle{fancy}

\begin{abstract}
Embodied artificial intelligence (AI) couples perception and learned decision making to actions that change the physical world. This coupling distinguishes an embodied agent from a conventional connected controller: the agent maintains task state and uncertainty, reasons about the consequences of actions, and adapts from subsequent observations. Wireless networking becomes relevant when perception, inference, or coordination is distributed, but it should not replace local safety control. This article develops a tutorial perception--communication--action (PCA) architecture that exposes task state, action deadlines, uncertainty, agent intent, and safety envelopes to a 6G orchestration plane. It separates capabilities already addressed by 5G and 5G-Advanced from functions that motivate 6G, including task-state interfaces, semantic freshness, predictive digital twins, and safety-aware coordination across agents. A multi-robot simulation study is retained to illustrate joint sensing, communication, and computation control. The results show where network orchestration improves task utility and where local autonomy remains essential.
\end{abstract}

\begin{IEEEkeywords}
Embodied AI, 6G networks, perception--communication--action loop, edge intelligence, semantic communication, multi-agent coordination.
\end{IEEEkeywords}

\section{Introduction \& Background}
\label{sec:intro}

Embodied AI refers to an agent whose intelligence is grounded in a body, a task, and repeated interaction with a physical environment. An embodied robot or vehicle does not only classify an observation. It estimates task state and uncertainty, selects an action subject to physical and safety constraints, observes the consequences, and updates its behavior. A remotely monitored sensor, a fixed industrial controller, or a digital twin is therefore not automatically an embodied AI agent. It becomes part of an embodied system only when it contributes to an adaptive closed loop whose actions alter subsequent observations~\cite{duan2022survey}.

This definition places embodied AI beside, but not inside, three established fields. Cyber physical systems emphasize verified interaction between computation and physical dynamics. Networked control studies stability under delay, loss, and quantization. Integrated sensing and communication and edge intelligence jointly allocate radio and compute resources. Embodied AI adds learned task state, world-model uncertainty, agent intent, adaptation, and the physical consequences of model errors. The novelty is thus not another sensing-to-control pipeline. It is an interface through which the network can serve these embodied variables without assuming responsibility for application-level control.

\begin{figure*}[t]
\centering
\includegraphics[width=0.9\textwidth]{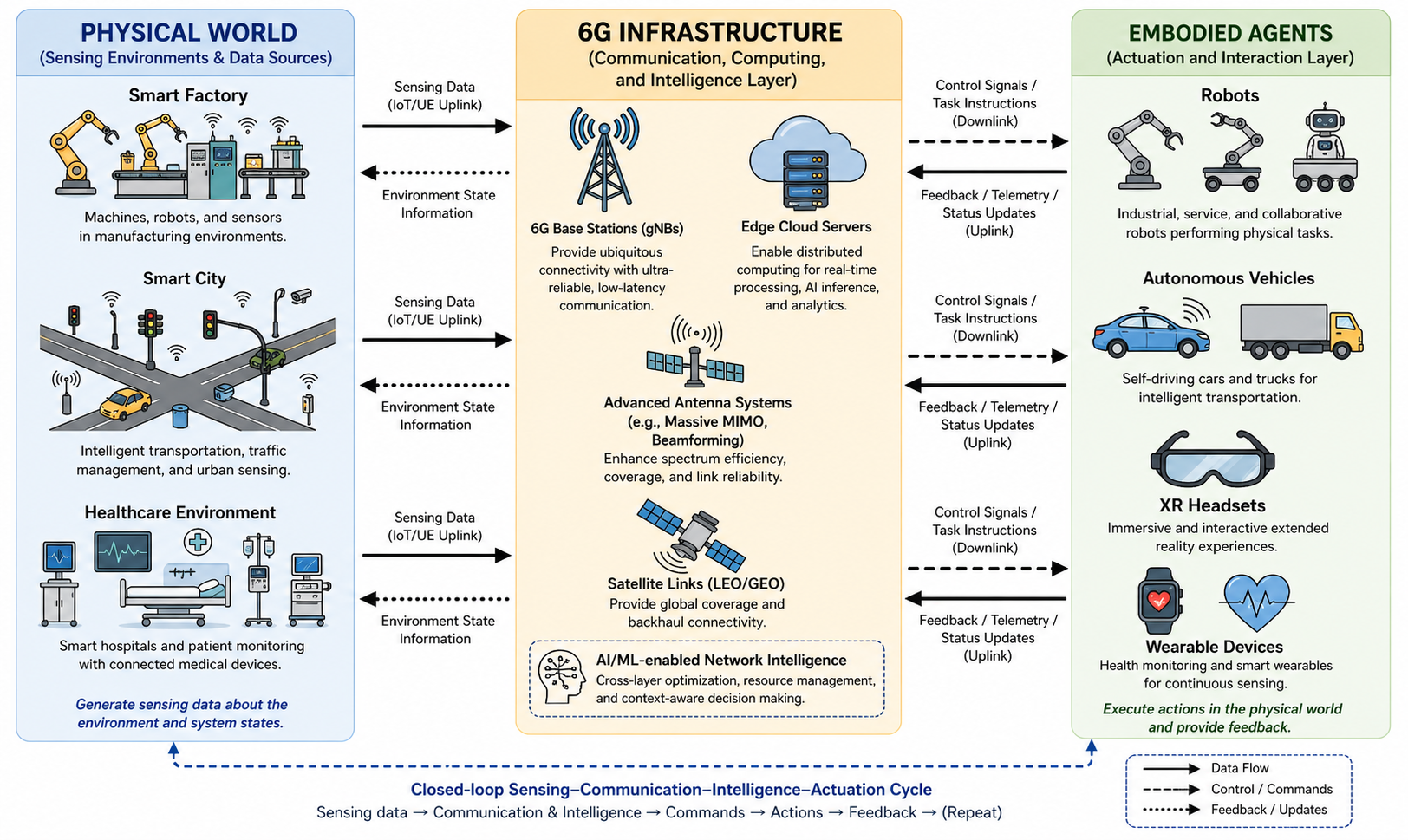}
\caption{Embodied intelligence over a wireless infrastructure. Only task-relevant state leaves an agent; safety-critical stabilization remains local. Edge and network functions assist perception, coordination, and prediction, while every action produces the next observation.}
\label{fig:system}
\end{figure*}

Fifth-generation systems already provide flexible numerologies, ultra-reliable low-latency communication, time-sensitive networking integration, private networks, positioning, and edge computing~\cite{itu2020m2150,3gpp22104, popovski2018control}. These capabilities are sufficient for many deterministic industrial loops. Sixth-generation networking is justified only when a task needs functions that are difficult to provide through local autonomy plus conventional connectivity: distributed multimodal perception, rapid exchange of learned scene representations, prediction of link and physical state, coordinated intent among many mobile agents, or joint adaptation of model placement and radio resources~\cite{strinati2019beyond}. Figure~\ref{fig:system} presents this bounded role. The network assists an embodied loop; it does not become the robot's uncertified controller.

This article makes three contributions. First, it defines the embodied information exposed at PCA interfaces: task state, uncertainty, deadline, intent, and safety envelope. Second, it maps those requirements to 5G baselines, candidate 6G enablers, and unresolved gaps. Third, it interprets a multi-robot resource orchestration study through physical-loop outcomes and safe fallback behavior. This framing separates AI for networking, networking for distributed AI, and networking for embodied agents.

\section{What Is New Beyond 5G and Conventional Closed Loops?}
Three roles of AI must not be conflated. \emph{AI for networking} uses learning for channel estimation, scheduling, or beam management; its output is a network configuration. \emph{Networking for distributed AI} transports training updates, activations, or inference requests; its objective is distributed model training and execution at the wireless edge~\cite{park2021semantic}. \emph{Networking for embodied agents} transports information whose age, uncertainty, and consistency affect a physical action. The first two may support the third, but neither is inherently embodied.

\begin{table*}[t]
\centering
\footnotesize
\caption{Representative loop requirements. Values are design examples, not universal 6G guarantees.}
\label{tab:requirements}
\begin{tabular}{p{2.2cm}p{2.3cm}p{2.5cm}p{3.0cm}p{5.0cm}}
\toprule
Use case & Relevant deadline & Traffic pattern & Deployment constraint & Interpretation \\
\midrule
Haptic teleoperation & Sub-millisecond to millisecond control update & Small bidirectional streams & Local edge and bounded jitter & A hard control deadline may apply to the local loop; wide-area assistance is predictive or supervisory~\cite{aijaz2019tactile}. \\
Cooperative driving & Tens of milliseconds, scenario dependent & Uplink-heavy features and intent & Mobility, occlusion, mixed operators & Tail age and cross-vehicle consistency matter more than average throughput~\cite{wang2020v2vnet}. \\
Mobile robots & Task and speed dependent & Bursty maps, features, and commands & Private edge, shared spectrum & Collision avoidance remains local; network assistance improves shared perception and planning. \\
Extended reality & Motion-to-photon deadline & Asymmetric pose and rendered content & Compute placement and user mobility & An application deadline is not identical to radio latency. \\
Digital twin assistance & Bounded state age & Periodic state plus event updates & Model fidelity and synchronization cost & Twin output is usable only while its uncertainty and state age remain inside an action-specific validity bound. \\
\bottomrule
\end{tabular}
\end{table*}

Table~\ref{tab:requirements} avoids presenting latency ranges as universal requirements. A hard deadline is a limit beyond which a controller may become unsafe or unstable. A tail constraint limits rare excessive delays. An average is useful for capacity planning but cannot certify a safety-critical loop. Each application must derive its bound from plant dynamics, controller design, and hazard analysis. The resulting 6G research problem is not simply lower latency. It is selective and accountable use of distributed intelligence. An agent should continue locally when a link fails, reject stale remote advice, and expose only the task state needed for coordination. Network support is most valuable above the fastest stabilization loop: cooperative perception, multi-agent planning, model update, and non-emergency trajectory optimization.

\section{Enablers and Remaining Gaps}
\begin{figure}[t]
\centering
\includegraphics[width=\columnwidth]{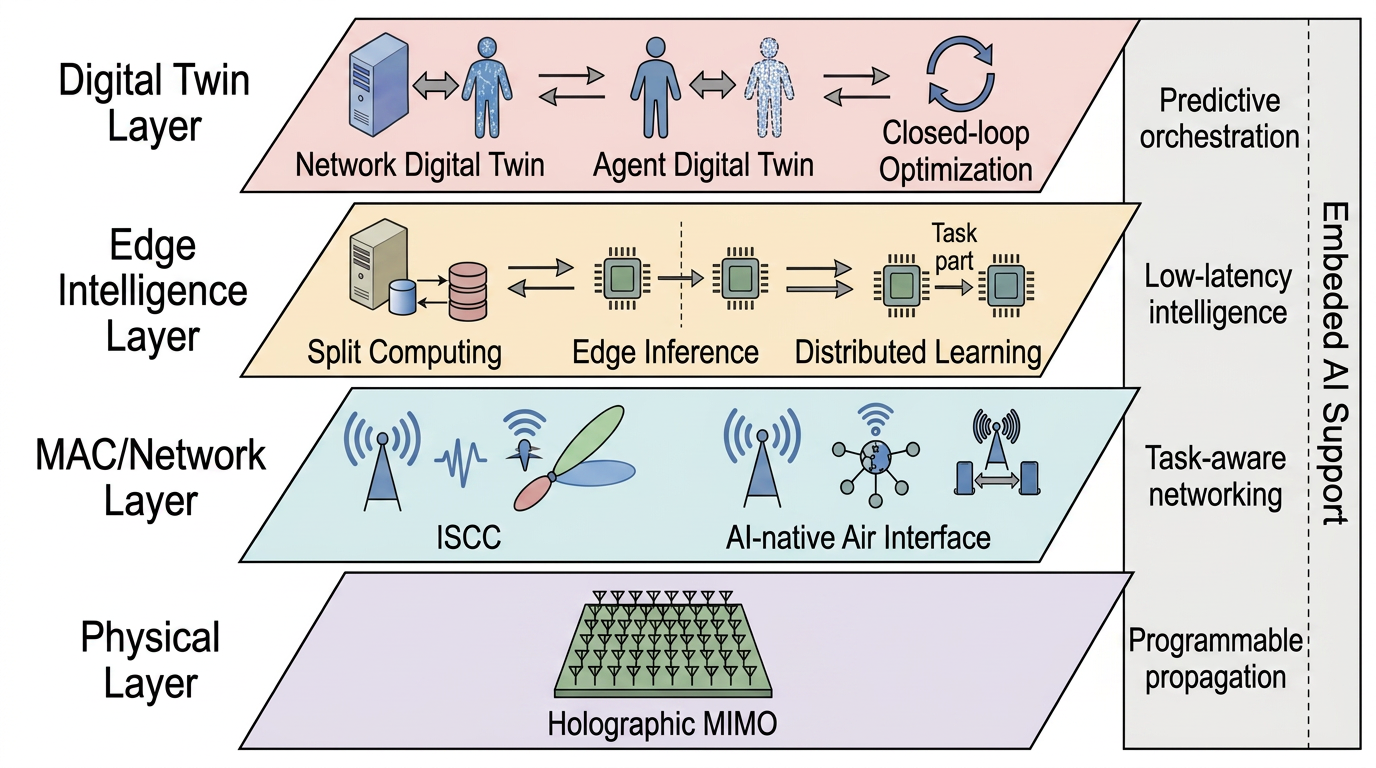}
\caption{Candidate 6G functions arranged by their contribution to embodied operation. Their value must be measured through task and safety outcomes, not by their presence in the protocol stack.}
\label{fig:stack}
\end{figure}

Figure~\ref{fig:stack} groups the main enablers. Integrated sensing, communication, and computation can supplement onboard perception with radio sensing and shared infrastructure~\cite{liu2022isac}. Semantic communication can transmit objects, latent features, or intent rather than reconstructing every source sample~\cite{xie2021deep}. Edge intelligence places time-sensitive inference near agents and reduces dependence on distant cloud infrastructure~\cite{satya2017edge,shi2016edge}. Predictive digital twins can test candidate actions and anticipate demand~\cite{faye2024twin}. Programmable propagation and AI-assisted air interfaces can improve coverage continuity, but these remain AI-for-networking functions unless their decisions are tied to a task state and safety policy.

\begin{table*}[t]
\centering
\footnotesize
\caption{Mapping from embodied requirements to network support and unresolved technical gaps.}
\label{tab:mapping}
\begin{tabular}{p{3.0cm}p{3.5cm}p{4.0cm}p{4.5cm}}
\toprule
Embodied requirement & Exposed interface information & Candidate support & Unresolved gap \\
\midrule
Timely action & Deadline class, state age, loop stage & Deadline-aware scheduling and edge placement & Joint tail guarantees across sensing, radio, queues, inference, and actuation \\
Task-relevant delivery & Task identifier, feature value, uncertainty & Semantic coding and adaptive sensing & Calibrated value metrics that remain valid after task or environment shift \\
Safe multi-agent action & Intent, priority, predicted occupancy, safety envelope & Cooperative perception and conflict-aware orchestration & Composable safety assurance across vendors and learned policies \\
Predictive operation & Twin age, fidelity, prediction uncertainty & Digital twin and proactive allocation & Validity tests and fallback when the twin is stale or inaccurate \\
Graceful degradation & Minimum local mode and link health & Local-edge-cloud split inference & Certified switching between assisted and autonomous modes \\
\bottomrule
\end{tabular}
\end{table*}

A concrete semantic metric is necessary. We use \emph{task gain per transmitted megabyte}: the improvement in task success, trajectory error, or safety margin obtained by sending a representation rather than using local information alone, divided by its communication volume. The numerator is selected by the application and evaluated against a local-only baseline. This metric prevents a compact representation from being called efficient when it saves bits but degrades control. Safety-critical messages are never discarded solely because their expected average gain is small.

\section{PCA Architecture and Interfaces}
\begin{figure*}[t]
\centering
\includegraphics[width=0.96\textwidth]{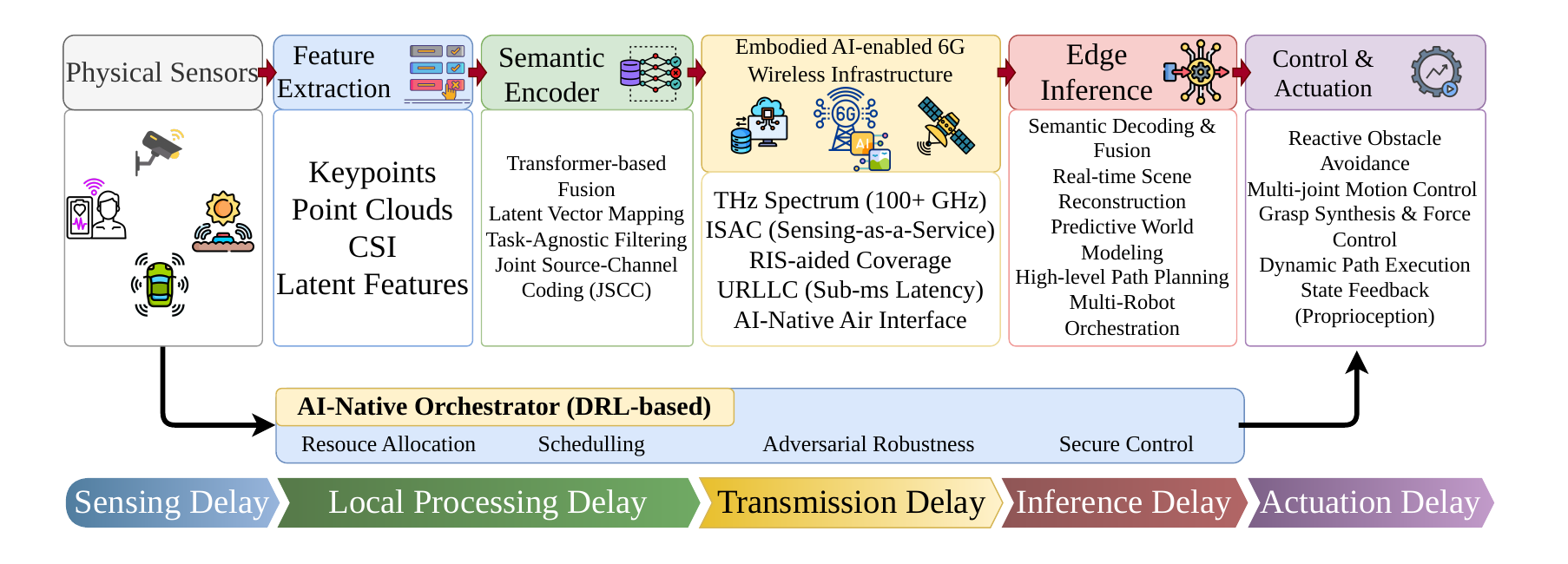}
\caption{PCA architecture. The forward interface carries task state, uncertainty, deadline, and intent. The return interface carries an action proposal, confidence, validity horizon, and required acknowledgment. A local safety monitor can reject or replace any remote proposal.}
\label{fig:pca}
\end{figure*}

Figure~\ref{fig:pca} decomposes one loop into sensing, local processing, transmission and queuing, edge inference, downlink delivery, and actuation. The sum of these stages is the loop latency. No equation is needed to expose the engineering implication: a radio improvement is useful only if it reduces the dominant stage and preserves enough margin before the action deadline. Timestamping at every boundary permits the orchestrator to attribute delay rather than treating communication delay as the entire loop.

\subsection{Forward and Return Interfaces}
At the forward interface, an agent sends a timestamped task-state message containing the task identifier, selected observations or features, calibrated uncertainty, state age, action deadline, predicted local trajectory, and a safety envelope. The message may also declare the value of an update relative to the agent's current local belief. At the return interface, an edge service sends an action proposal or shared world-state update with confidence, a validity horizon, provenance, and the state version used for inference. The agent acknowledges acceptance or reports rejection and the locally executed fallback. This feedback lets the network learn whether delivery changed physical behavior. Timing is event driven for hazards and periodic for state maintenance. Fast stabilization executes on the device. Edge inference serves a slower planning loop, and the cloud supports model training and fleet-level analysis. This hierarchy avoids placing a variable wide-area delay inside a loop that requires deterministic local timing.

\subsection{Conflict Resolution and Safety}
A shared environment can produce incompatible action proposals. We use a four-level arbitration rule: human and emergency-stop commands have highest priority; certified collision-avoidance actions follow; mission-critical coordinated actions are next; efficiency and comfort actions are lowest. Within a level, the orchestrator admits only proposals whose predicted occupancy does not violate another agent's safety envelope. If agreement is not reached before the deadline, each agent enters a predefined minimum-risk condition, such as braking, hovering, or reducing speed. A learned policy may rank feasible actions but cannot override the certified safety monitor.

\subsection{Digital Twin Validity}
A twin update contains its timestamp, synchronization interval, model version, and prediction uncertainty. Before actuation, the agent checks whether the twin age and uncertainty satisfy an action-specific validity rule. Slow planning can tolerate older state than collision avoidance. A failed check discards the twin output and invokes local sensing or conservative control. Synchronization should therefore be adaptive: event-triggered near hazards or rapid motion and less frequent during steady operation. This makes twin fidelity an explicit control input rather than a broad promise of predictive networking.

\section{Multi-Robot Simulation Study}
\begin{figure*}[t]
\centering
\includegraphics[width=0.98\textwidth]{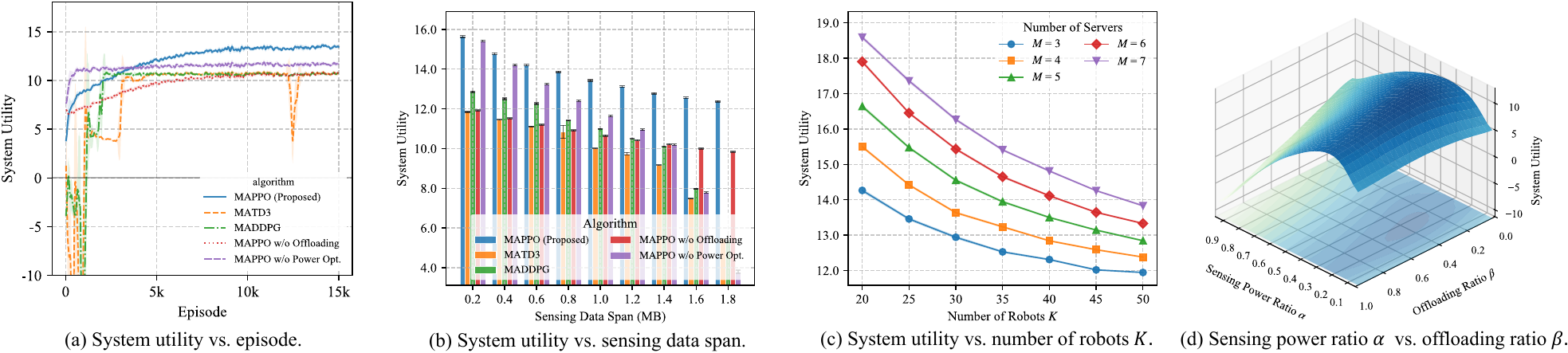}
\caption{Multi-robot results retained from the submitted study: training convergence; utility versus sensing payload; utility versus robot and server counts; and the sensing-power/offloading operating surface. Utility aggregates task accuracy, loop delay, and energy penalties.}
\label{fig:results}
\end{figure*}

The study considers mobile robots that generate visual and LiDAR features and share wireless and edge-compute capacity. At each decision interval, every robot selects a sensing power level, an offloading fraction, and communication and compute resources. The observable state includes local channel quality, queue and task size, residual energy, available server capacity, and the age of the latest task state. Actions are coupled because robots contend for spectrum and servers. The reward increases with perception task accuracy and timely task completion and decreases with loop delay, energy use, and constraint violations. A collision-risk or trajectory-error penalty provides the physical-loop link; unsafe actions are masked by the local safety rule described above. 

The evaluation varies sensing payload from 0.2 to 1.8 MB, the robot population from 20 to 50, and the number of edge servers from three to seven. Training runs for 15,000 episodes. MAPPO~\cite{yu2022mappo} is compared with MATD3, MADDPG, a no-offloading ablation, and a no-sensing-power-optimization ablation. All methods receive the same observations, action limits, channel samples, task arrivals, and compute budgets. The curves in Fig.~\ref{fig:results} should be interpreted as illustrative normalized utility, not a universal 6G performance claim. For reproducibility, the submission package should additionally archive the exact channel model, bandwidth and power limits, processor rates, random seeds, reward weights, network sizes, learning rates, discount factor, clipping value, batch size, and update schedule used to generate the retained plots. These numerical settings were not contained in the supplied manuscript source and must be verified against the authors' simulation code before submission.

Figure~\ref{fig:results}(a) shows that MAPPO reaches the highest stable utility after early exploration. Figure~\ref{fig:results}(b) shows degradation as larger sensing payloads increase radio and compute demand; the proposed policy retains the largest margin. Figure~\ref{fig:results}(c) separates congestion from added edge capacity: more robots reduce utility, whereas more servers partly restore it. Figure~\ref{fig:results}(d) shows that neither maximum sensing power nor maximum offloading is optimal. The operating point balances perception quality against transmission, queueing, and compute cost. The study supports a limited conclusion: joint orchestration can improve an embodied task when perception quality and physical-loop penalties are included. It does not show that 6G should control robots directly. A stronger validation should report tail loop latency, task completion, trajectory error, collision or safety-envelope violations, energy, and performance during link loss. Local-only, communication-only, and oracle full-information baselines would reveal when network assistance is genuinely beneficial.

\section{Deployment and Research Directions}
Three challenges determine whether this architecture is practical. First, task-state semantics must be interoperable. A scheduler cannot use uncertainty, deadline, or intent if each robot encodes them differently. Standardization should define transportable descriptors and versioning, not prescribe application policies. IEEE work on AI integration in mobile networks provides a starting point~\cite{ieeep1948}. Second, safety evidence must cross subsystem boundaries. Radio reliability alone cannot certify a robot, while end-to-end certification cannot ignore queues, model drift, or stale shared perception. Digital twin and hardware-in-the-loop testbeds should reproduce channel dynamics, physical motion, inference faults, and fallback transitions. Reports should include tail distributions and hazard exposure, not only mean utility. Third, deployment must be incremental. Private 5G, local edge servers, and onboard autonomy can support early systems. Candidate 6G functions should be added only where measured task gain exceeds integration, energy, privacy, and certification costs. Raw sensor streams should remain local when possible; task features and intent should be authenticated, access controlled, and retained only for a defined purpose.

\section{Conclusion}
Embodied AI does not require a network to solve application control. It requires the network to understand when distributed information is useful, stale, uncertain, or unsafe. The proposed PCA architecture makes this boundary explicit through task-state and action interfaces, hierarchical timing, conflict arbitration, twin-validity checks, and local fallback. This framing identifies a credible 6G role beyond existing 5G connectivity: supporting shared perception and coordinated intelligence while preserving certified local autonomy. The retained multi-robot study illustrates the resource tradeoff, but future evidence must center on physical outcomes and tail behavior.

\balance
\bibliographystyle{IEEEtran}
\bibliography{ref}

\end{document}